\documentclass[prd,twocolumn,unsortedaddress,superscriptaddress,a4paper,nofootinbib,%
nobalancelastpage,preprintnumbers,showpacs]{revtex4}
\usepackage{longtable}
\usepackage{latexsym,amsbsy}
\usepackage[pdftex]{graphicx}
\newcommand{\sutf}{\ensuremath{\mathrm{SU(3)_F}}}
\newcommand{\DD}{\ensuremath{\mathrm{D}\overline{\mathrm{D}}{}^*}}
\newcommand{\be}{\begin{equation}}
\newcommand{\ee}{\end{equation}}
\newcommand{\SpSp}[2]{ \mbox{$\vec{\sigma }_{#1}.\vec{\sigma }_{#2}$}}
\newcommand{\lala}[2]{ \mbox{${\tilde{\lambda}_{#1}\cdot
         \tilde{\lambda}_{#2}}$}}
\renewcommand{\vec}[1]{\boldsymbol #1}
\newcommand{\df}[2]{\ensuremath{ {\raise
1pt\hbox{$\displaystyle #1$}\over \raise -2pt \hbox{$\displaystyle
#2$}}}}
\begin{document}
\title{\bf A chromomagnetic mechanism for the X(3872) resonance} 
\date{\today}
\author{H.~H\o gaasen}
\email{hallstein.hogasen@fys.uio.no}
\affiliation{Department of Physics, University of Oslo,
Box 1048  NO-0316 Oslo Norway}
\author{J.-M.~Richard}
\email{jean-marc.richard@lpsc.in2p3.fr}
\affiliation{Laboratoire de Physique Subatomique et Cosmologie, 
 Universit\'e Joseph Fourier--IN2P3-CNRS\\[-1pt]
53, avenue des Martyrs, 38026  Grenoble cedex, France}
\author{P.~Sorba}
\email{sorba@lapp.in2p3.fr}
\affiliation{Laboratoire d'Annecy-le-Vieux de Physique Th\'eorique
(LAPTH)\\[-1pt]
9 chemin de Bellevue, B.P. 110, 74941 Annecy-le-Vieux Cedex, France}
\preprint{\begin{minipage}{3cm}LAPTH-1221/05\\
LPSC-05-112\\
hep-ph/0511039\end{minipage}}
\pacs{12.39.-x,12.39.Mk,12.40.Yx}
\begin{abstract}
The chromomagnetic interaction, with proper account for
flavour-symmetry breaking, is shown to explain the mass and coupling
properties of the X(3872) resonance as a $J^{PC}=1^{++}$ state
consisting of a heavy quark--antiquark pair and a light one.  It is
crucial to introduce all the spin--colour configurations compatible
with these quantum numbers and diagonalise the chromomagnetic
interaction in this basis.  This approach thus differs from the
molecular picture \DD\ and from the diquark--antidiquark picture.
\end{abstract}
\maketitle
%
%
In recent months, several intriguing new hadron states have been
announced.  Some of them are rather controversial since tentatively
seen in some experiments and not in others.  On the other hand, the
X(3872) reported by the Belle collaboration \cite{Choi:2003ue} can be
considered as well established, since it has been confirmed at BaBar
and at the Fermilab collider
\cite{Abazov:2004kp
}. There are some indications that
it could have $J^{PC}=1^{++}$ quantum numbers \cite{Abe:2004sd}, in
particular it is not seen in a $\gamma\gamma$ search at CLEO
\cite{Dobbs:2004di}.

Several theoretical explanations have been proposed for the X(3872).
It could be mainly a charmonium excitation $(c\bar{c})$, though none
of the partial-wave assignments ${}^{2s+1}\mathrm{L}_J$ actually
matches the predictions of charmonium models tuned to fit the known
levels \cite{Abe:2004sd,Barnes:2003vb
}.

An hybrid scenario has also been suggested for this state, or for the
other states discovered in this region: X(3940) \cite{Abe:2004zs} or
Y(4260) \cite{Aubert:2005rm}.  Excitations of the string linking the
quark to the antiquark, or, in the QCD language, of the gluon field,
were proposed long ago on the basis of
some models 
and confirmed by Lattice QCD
.  A signature of this
would be a decay with at least one orbitally excited meson, for
instance $\mathrm{D}^{**}+ \overline{\mathrm{D}}$ \cite{Kou:2005gt}.

The Yukawa mechanism is not
restricted to the nucleon--nucleon system, and holds for any pair of
hadrons containing light quarks.  In particular, pion-exchange, if
allowed and attractive, can be just strong enough to bind heavy
hadrons to form a deuteron-like compound. 
Remarkably, this mechanism led some authors to predict the existence
of $\DD+ \mathrm{c.c.} $ states and when the X(3872) was found very
close to the \DD\ threshold, it was considered as a very natural
candidate 
\cite{Swanson:2003tb
}.  However, some uncertainties remain: though the pion-coupling is
deduced from the nucleon--nucleon case, the
$\mathrm{D}\mathrm{D}^*\pi$ form factor is not known accurately as
well as the short-range part of the interaction needed to supplement
pion exchange. Also, due the mass difference between $\mathrm{D}$ and $\mathrm{D}^*$, the Yukawa potential might be of shorter range in $\DD$ than in the nucleon--nucleon case, and hence be less effective \cite{Suzuki:2005ha}.

More generally, several approaches are based on the X(3872) having
mainly a $(cq\bar{c}\bar{q})$ quark content, where $q$ denotes a light
quark.  Besides the nuclear-physics approach, schematically noted
$(c\bar{q})-(\bar{c}q)$, an interesting come-back of the diquark
concept has been observed in the recent literature.  In particular,
Maiani et al.\ proposed to describe simultaneously the X(3872) and
X(3940) as $(cq)(\bar{c}\bar{q})$ states, and Y(4260) as a
$(cs)-(\bar{c}\bar{s})$ state with an orbital momentum $\ell=1$
between the diquark and the antidiquark
\cite{Maiani:2004vq
}.  This is a rather elegant
picture, but the mass of the diquark is not known and has to be
adjusted empirically.

None of the available models has won an overall consensus, yet, and
the door remains open for another binding mechanism.  This is the aim
of the present letter.  More details, as well as applications to other
spin-flavour combinations will be presented in a forthcoming article.
The starting point is the chromomagnetic interaction, inspired by the
one-gluon-exchange contribution \cite{DeRujula:1975ge}, but covering a
wider class of model with a spin--spin interaction that bears the
colour dependence of a colour-octet exchange.  The chromomagnetic
interaction gives a convincing explanation of the mass splittings of
ordinary hadrons and has been decisive in promoting the possibility of
hadron states with a multiquark content.  In particular, some S-wave
$(q^2\bar{q}^2)$ states can well be lighter than the P-wave
excitations of the $(q\bar{q})$ system, to explain why supernumerary
scalar states are observed with a low mass \cite{Jaffe:1976ig}.
Exotic configurations can also occur, due to a coherent chromomagnetic
attraction that is larger than the sum of chromomagnetic effects in
the decay products~\cite{Jaffe:1976yi
}.

In pioneering papers on chromomagnetic effects applied to multiquark
states, ordinary ($q=u,d$) and strange ($s$) quarks were treated in
the limit of $\sutf$ flavour symmetry.  However, when its breaking is
introduced, the chromomagnetic attraction of the $\Lambda$ baryon is
not changed, while that of $\mathrm{H}=(ssuudd)$ decreases.  Hence the
stability of the H-dibaryon is weakened by $\sutf$ breaking.  A
similar effect is observed for the 1987-vintage pentaquark,
$\mathrm{P}=(\overline{Q}sqqq)$.  See, e.g.,
Ref.~\cite{Rosner:1985yh
}.  Another difficulty is that the strength of the chromomagnetic
force, related to the quark--quark short-range correlation, is
probably smaller in H or P than in ordinary hadrons.  It thus seems
necessary to refine the treatment of chromomagnetic effects.

The present study takes full account of flavour symmetry breaking when
estimating the chromomagnetic interaction of multiquarks, and it
happens that this treatment provides a very good candidate for the
X(3872), with about the right mass, and the right coupling patterns,
namely \DD\ and $\mathrm{J}/\psi$ plus a light vector meson.

%
The interaction Hamiltonian acting on the colour and spin degrees of
freedom reads
\begin{equation}\label{eq:Hcm}
H_{\mathrm{CM}} = - \sum_{i,j} C_{ij}\,\lala{i}{j}\,\SpSp{i}{j},
\end{equation}
where the coefficients $C_{ij}$  depend on the quark masses and
properties of the spatial wave function. In absence of a complete
theory, this Hamiltonian leads to a mass formula
\begin{equation}\label{eq:Mcm}
\mathcal{M} =\sum_i m_i -\langle
\sum_{i,j}C_{ij}\,\lala{i}{j}\,\SpSp{i}{j}\rangle,
\end{equation}
with effective masses $m_i$ which include constituent masses and their
chromoelectric energy (binding effect).  This formula reflects the
basic symmetry principles which govern the ground-state hadron masses.
The solution of the eigenvalue problem for the chromomagnetic term is
thus of interest, not only in spectroscopy, but in all the reactions
where a quark or an antiquark interacts with a system of other quarks,
for instance, final-state interaction in weak decays.

A useful phenomenology can be developed on the basis of mass formulae
such as (\ref{eq:Mcm}).  See, for instance,
Ref.~\cite{Lichtenberg:1995kg}.  The mesons being more tightly bound
than baryons, the fits usually lead to lighter values of the effective
masses $m_i$ for mesons, and larger correlation coefficients.  This is
in qualitative agreement with model calculations which can be
performed within the harmonic oscillator model, or with more general
inequalities relating mesons to baryons \cite{Nussinov:1999sx}.  A fit
within $\pm 10\;$ MeV of charmed baryons gives the set of masses
\begin{equation}\label{eq:masses}
m_c=1550\,\mathrm{MeV},\  m_q=450\,\mathrm{MeV},\
m_s=590\,\mathrm{MeV},
\end{equation}
and  strength factors 
\begin{equation}\label{eq:par}
\begin{array}{lll}
  C_{qq}=20\,\mathrm{MeV},&
C_{qc}=5\,\mathrm{MeV},&C_{qs}=15\,\mathrm{MeV},\\
C_{ss}=10\,\mathrm{MeV},&
C_{cs}=4\,\mathrm{MeV}, &C_{c\overline{c}}=4\,\mathrm{MeV}.
\end{array}\end{equation}

For $(q\bar{q})$ mesons, $H_\mathrm{CM}=16\,C_{12} \SpSp{1}{2}/3$, and
for $(q_1q_2q_3)$ baryons with three valence quarks, $\langle
\lala{i}{j}\rangle=-8/3$ factors out for all pairs.  Estimation of the
value of $\sum \lala{i}{j}\,\SpSp{i}{j}$ for more complicated systems,
once an overall strength has been factored out (i.e., in the flavour
symmetric limit), has been carried out in \cite{Jaffe:1976ih} and
further developed by several authors.  The formulae involve the
Casimir operators of SU(2) (spin), SU(3) (colour or flavour) and SU(6)
(spin--colour).  As we are dealing with states combining heavy and
light quarks, and even account for \sutf\ breaking in the light
sector, we cannot assume that all $C_{ij}$ are equal.  Hence for any
given $J^{PC}$ set of quantum numbers, we list all possible
colour--spin states and write down explicitely $H_\mathrm{CM}$ in this
basis.

In the case of colour-singlet, $J^{PC}=1^{++}$, a basis can be built
with (1,3) and (2,4) subsystems having a well defined colour
(superscript 1 for singlet and 8 for octet) and spin (0 or 1 in
subscript)
\begin{eqnarray}\label{eq:alphai}
\hskip -10pt& \alpha_1 =(q_1 \overline{q}_3)^1_0 \otimes (q_2
\overline{q}_4 )^{{1}}_{1},\ 
&  \alpha_2 =(q_1 \overline{q}_3)^1_1 \otimes (q_2 \overline{q}_4
)^{{1}}_{0},\nonumber\\
\hskip -10pt& \alpha_3 =(q_1 \overline{q}_3)^1_1 \otimes (q_2
\overline{q}_4 )^{{1}}_{1},\ 
&\alpha_4 =(q_1 \overline{q}_3)^8_0 \otimes (q_2 \overline{q}_4
)^{{8}}_{1}, \\ 
\hskip -10pt& \alpha_5 =(q_1 \overline{q}_3)^8_1 \otimes (q_2
\overline{q}_4 )^{{8}}_{0}, \ 
& \alpha_6 =(q_1 \overline{q}_3)^8_1 \otimes  (q_2 \overline{q}_4
)^{{8}}_{1} . \nonumber
\end{eqnarray}

\begin{widetext}
\begin {longtable}{c}
\caption{\label{Tab1}Colourmagnetic Hamiltonian $-H_\mathrm{CM}$ in
the basis (\ref{eq:alphai})}
\end{longtable}
\vskip -20pt
$$\left[\vbox{\hskip -10pt
\begin {tabular}{cccccc}
$16C_{13}-\df{16}{3}C_{24}$&0&0&0&$\df{8\sqrt2}{3}(C_{23}+C_{12})$&0\\
0&$-
\df{16}{3}C_{13}+16C_{24}$&0&$\df{8\sqrt2}{3}(C_{23}+C_{12})$&0&0\\
0&0&$-\df{16}{3}(C_{13}+C_{24})$&0&0&$\df{8\sqrt2}{3}(C_{23}-C_{12})$\\
0&$\df{8\sqrt2}{3}(C_{23}+C_{12})$&0&$\df{2}{3}C_{24}-2C_{13}$&
$\df{28}{3}C_{23}-\df{8}{3}C_{12}$&0\\
$\df{8\sqrt2}{3}(C_{23}+C_{12})$&0
&0&$\df{28}{3}C_{23}-\df{8}{3}C_{12}$&
$-2 C_{24}+ \df{2}{3}C_{13}$&0\\
0&0&$\df{8\sqrt2}{3}(C_{23}-C_{12})$
&0&0&$\df{2}{3}(4C_{12}+14C_{23}+C_{13}+C_{24})$
\end{tabular}}\ \right]$$
\end{widetext}

For $(cq\bar{c}\bar{q})=(1,2,3,4)$ states, the calculation is
simplified since $C_{14}=C_{23}$ and $C_{12}=C_{34}$ by charge
conjugation symmetry.\footnote{The first calculation relevant for this
case was made by G.\ Gelmini \protect\cite{Gelmini:1979ir}.} The
Hamiltonian $-H_\mathrm{CM}$ acting on the basis (\ref{eq:alphai}) is
represented by the matrix given in Table~\ref{Tab1}.

It is immediately seen from this matrix that in the case where the
chromomagnetic interaction is the same for a quark--quark pair as for
the quark--antiquark pair, i.e., $C_{12}=C_{23}$, there is an
eigenvector with eigenvalue $-(8C_{12}+28 C_{23}+ 2 C_{13}+ 2C_{24})
/3$ for the colourmagnetic Hamiltonian, which is a pure colour octet\
$\otimes$\ octet, spin $(s=1)\otimes(s=1)$ state, $\alpha_6=(c
\overline{c})^8_1 \otimes ( q \overline{q})^{{8}}_{1}$.  This state
therefore cannot freely dissociate into a charmonium state and a light
meson!

This eigenstate of the chromomagnetic Hamiltonian can freely fall
apart in two mesons carrying charm.  However, if the state is
rewritten in the (1,4)(2,3) basis, corresponding to charmed mesons,
i.e., $(c\bar{q})(q\bar{c})$,
\begin{eqnarray}\label{eq:betai}
\hskip -10pt&\beta_1 =(q_1 \overline{q}_4)^1_0 \otimes (q_2
\overline{q}_3 )^{{1}}_{1},\ 
&\beta_2 =(q_1 \overline{q}_4)^1_1 \otimes (q_2 \overline{q}_3
)^{{1}}_{0}, \nonumber \\
\hskip -10pt&\beta_3 =(q_1 \overline{q}_4)^1_1 \otimes (q_2
\overline{q}_3 )^{{1}}_{1}, \ 
& \beta_4 =(q_1 \overline{q}_4)^8_0 \otimes  (q_2 \overline{q}_3
)^{{8}}_{1}, \\ 
\hskip -10pt&\beta_5 =(q_1 \overline{q}_4)^8_1 \otimes  (q_2
\overline{q}_3 )^{{8}}_{0}, \ 
&\beta_6 =(q_1 \overline{q}_4)^8_1 \otimes (q_2  \overline{q}_3
)^{{8}}_{1}, \nonumber
\end{eqnarray}
using the crossing matrix from the basis in Eq.\ (\ref{eq:alphai})
to the basis in Eq.\ (\ref{eq:betai}),
\begin{equation}\label{eq:cross1}
{ \renewcommand{\arraystretch}{1.6}
\left[ %
\begin {array}{cccccc}
\df{1}{6}&\df{1}{6}&\df{1}{3\sqrt2}&\df{2}{3\sqrt2}&\df{2}{3\sqrt2}&\df{2}{3}\\
\df{1}{6}&\df{1}{6}&-\df{1}{3\sqrt2}&\df{2}{3\sqrt2}&\df{2}{3\sqrt2}&-\df{2}{3}\\
\df{1}{3\sqrt2}&-\df{1}{3\sqrt2}&0&\df{2}{3}&-\df{2}{3}&0\\
\df{2}{3\sqrt2}&\df{2}{3\sqrt2}&\df{2}{3}&-\df{1}{6}&-\df{1}{6}&-\df{1}{3\sqrt2}\\
\df{2}{3\sqrt2}&\df{2}{3\sqrt2}&-\df{2}{3}&-\df{1}{6}&-\df{1}{6}&\df{1}{3\sqrt2}\\
\df{2}{3}&-\df{2}{3}&0 &-\df{1}{3\sqrt2}&\df{1}{3\sqrt2}&0
\end {array} \right],}
\end{equation}
it is immediately realised that there is little octet--octet content
for this eigentstate in this crossed basis, and that the coulour
singlet--singlet part just corresponds to the charmed mesons D and
$\overline{\mathrm{D}}{}^*$ (or c.c.).  Due to the lack of phase
space, this decay is strongly suppressed.

If the condition $C_{12}=C_{23}$ is relaxed and a different
interaction strength is allowed for the quark--quark and
quark--antiquark pairs, the interesting eigenvector of the
chromomagnetic Hamiltonian is readily seen to acquire a small
component on the state $ \alpha_3 $, but not on $ \alpha_1$ or on
$\alpha_2 $.  This means that this eigenstate of $H_{\mathrm{CM}}$
will choose to disintegrate into a $\mathrm{J}/\psi$ and an ordinary
vector meson, just as the X(3872) does.  There is no amplitude for
dissociation into charmonium and a pseudoscalar meson, at least at the
level of the mere quark rearrangement.

Instead of an analytical proof which involves some tedious $6\times 6$
linear algebra, a numerical illustration will be given.  If the
parameters (\ref{eq:par}) are adopted and if $C_{\bar{q}c}=C_{qc}$ is
further assumed, the eigenstate $\alpha_6$ receives a chromomagnetic
energy $-76\,$MeV. If a value $C_{\bar{q}c}=6.5\,$MeV is adopted
instead, an eigenvector $\sum_i a_i \alpha_i$ is obtained, with
\begin{equation}\label{eq:eigenv1}
\{a_i\}=\{0,0,\epsilon=0.026,0,0,\sqrt{1-\epsilon^2}\},
\end{equation}
i.e., a very small $\mathrm{J}/\psi +\rho$ or $\mathrm{J}/\psi
+\omega$ component, and its eigenvalue is now $-90\,$MeV. If inserted
in Eq.\ (\ref{eq:Mcm}), it corresponds to a mass
$\mathcal{M}(X)=3910\;$MeV with the parameters (\ref{eq:masses}),
close to the observed mass $3872\;$ MeV. Several corrections can be
anticipated, for instance a coupling to the \DD\ channel.

It is worth stressing that the above state has \emph{not} the lowest
eigenvalue for $H_{\mathrm{CM}}$.  Another eigenstate exists with a
much lower eigenvalue, about $-220\,$MeV. This state, $\sum_i b_i
\alpha_i$ with
\begin{equation}\label{eq:eigenv2}
\{b_i\}=\{ - 0.0026,- 0.989,0,- 0.146,- 0.021,0.0\},
\end{equation}
is seen to be almost completely coupled to the channel consisting of
$\mathrm{J}/\psi$ and a light pseudoscalar and therefore is probably
very broad and is just a part of the continuum.

The remarkable eigenstate of the chromomagnetic Hamiltonian actually
consists of four states, namely $ X_{+}=(cu\bar{c}\bar{d})$,
$X_{-}=(cd\bar{c}\bar{u})$, $Y_{1}=(cu\bar{c}\bar{u})$ and
$Y_{2}=(cd\bar{c}\bar{d})$.  They all receive a contribution from the
QCD version of the Pirenne annihilation
 potential \cite{Pirenne:1947
} acting on $(c\bar{c})$. In addition, the two neutral states mix
through
annihilation of the $(u\bar{u})$ and $(d\bar{d})$, colour octet,
spin~1, components\footnote{Mixing with glueballs, hybrids and
high-mass tetraquark states is neglected}.

With this mixing and the mass difference between $u$ and $d$ quarks,
the isospin zero state $(Y_{1}+Y_{2})$ and the neutral isospin one
state $(Y_{1}-Y_{2})$ are not anymore eigenstates of the Hamiltonian.
The mass matrix governing the physical states is
\begin{equation}\label{eq:mixing}
  \left[ \begin {array}{cc} \ -a&-a\\
  -a& 2(m_d-m_u)-a\end {array} \right],
\end{equation}
where $a$ is the annihilation potential term.

In the one-gluon-exchange model with free gluons in the intermediate
state, the strength $C_{q\bar{q}}$ of $t$-channel exchange and $a$ of
$s$-channel exchange are related by $a=6C_{q\bar{q}}$.  However,
perturbation theory with confined gluons suggests $a\simeq
C_{q\bar{q}}$ \cite{Gelmini:1979ir}.  If a value $a=15\;$MeV is taken
for the annihilation term and $m_d-m_u=3.5\;$MeV, the ``mostly $I=1$''
state lies $31\;$MeV above the ``mostly $I=0$'' state. The lowest state, mostly $I=0$,  has an amplitude for
$\mathrm{J}/\psi+\rho$ decay which is about 0.11 times the amplitude
for $\mathrm{J}/\psi+\omega$ decay: this is roughly what is needed to
explain the branching ratio of X(3872) for the two different final
states. The observed branching ratios are about equal
although phase space strongly favours $\mathrm{J}/\psi+\rho$ decay, since 
only the low-mass tail of the $\omega$ is kinematically allowed (see \cite{Abulencia:2005zc} for a more detailed discussion on this point).  A
further shift of the $I=0$ and $I=1$ states is induced by the nuclear
forces acting on the long-range $(c\bar{q})(q\bar{c})$ part of the
wave function, favouring $I=0$.  The effect is there, even if this is
not the main binding mechanism in our approach. The state with mostly $I=1$ isospin content, should be seen as a broad resonance decaying into  into \DD\  or $\mathrm{J}/\psi+\pi\pi$.

It is natural to ask what happens if the flavour content of
$(cq\bar{c}\bar{q})$ is modified, while keeping the $J^{PC}=1^{++}$
quantum numbers.  For $(qq\bar{q}\bar{q})$ and $(sq\bar{s}\bar{q})$,
the state is well above the threshold of two mesons.  This appears
also to be the case for the $(cs\bar{c}\bar{s})$, $(bs\bar{b}\bar{s})$
and $(bc\bar{b}\bar{c})$ configurations.  (For this last
configuration, since the spin excitation of the $\mathrm{B}_c$ meson
is not known, it has been neccesary to extract the value the
coefficient $C_{b\bar{c}}$ from theoretical calculations
\cite{Eichten:1994gt}.)\@ The situation is different for the
$(bq\bar{b}\bar{q})$ states: if the parameters are tuned to fit the
the measured values of the masses of B, B${}^*$, $\Upsilon$ and
$\Lambda_b$ hadrons, the $(bq\bar{b}\bar{q})$ states appears as stable
against dissociation into $\mathrm{B}\overline{\mathrm{B}}{}^*$.  It can decay into $\Upsilon+\omega$.
%
%
To summarise, the chromomagnetic interaction, acting on the
configurations $(cq\bar{c}\bar{q})$ with hidden charm, has been shown
to single out a remarkable state which is an almost pure octet--octet
state in the $(c\bar{c})+(q\bar{q})$ channel.  It has a large
singlet--singlet component of the type
$\mathrm{D}+\overline{\mathrm{D}}{}^*$ in the crossed $(c\bar{q})+
(\bar{c}q)$ channel.  However, this decay is kinematically strongly
suppressed, as the state is at about the same mass as this threshold.
A small impurity gives a small branching ratio into
$\mathrm{J}/\psi+\rho$ and $\mathrm{J}/\psi+\omega$, the former being
favoured by phase space, whilst $\mathrm{J}/\psi+\pi$ is suppressed.
This hadron is thus rather narrow, a remarkable property for a
multiquark without internal orbital momentum between clusters.  This
state is therefore a most natural candidate for describing the
X(3872).

Since the time of baryonium ``colour chemists'' thought that colour
will show up as a new spectroscopic degree of freedom, and states such
as ``mock-baryonium'', ``meso-baryons'' or ``pseudomesonium'' were
proposed, with colour-triplets, sextets or octets at both ends of a
rotating colourelectric string
\cite{Johnson:1975sg
 }.   However, it
was never convincingly explained how such a clustering could occur
from the dynamics of confinement.  Our state should be more easily accepted, 
since the two quarks and the two antiquarks are in an overall
S-wave.

Further measurements of the properties of the X(3872)
will help to test the chromomagnetic mechanism, which
furthermore predicts other interesting states, especially in
configurations combining heavy and light flavours. An example is $(bc\bar{q}\bar{q})$ with $J^P=1^+$. This will be
studied in a forthcoming paper.  It is simply stressed here that the
mechanism proposed for the X(3872) requires very specific spin and
flavour configurations.  This explains why multiquark states are so
elusive in the hadron spectrum

 We benefitted from interesting discussions with F.\ Buccella.
 Comments on the mansucript by M.\ Asghar are gratefully acknowledged.
 One of us (H.\ H.) would like to thanks LAPTH, Annecy-le-Vieux, for
 the hospitality extended to him.
%

%
\end{document}